\documentstyle[twoside,fleqn,espcrc2]{article}

\input epsf.sty
\newdimen\psfigsize
\def\psfigure#1 #2 #3 #4 #5{
    \begin{figure}[tbh]
    \vbox{
    \null\hskip#2\epsfxsize=#1 \epsfbox[0 0 4096 4096]{#4}
    \vskip -0.3in
    \caption {#5 \label{#3}}
    \vskip -0.2truein plus0.2truein}
    \end{figure}
}
\def\psoddfigure#1 #2 #3 #4 #5 #6{
    \begin{figure}[tbh]
    \vbox{
    \null\hskip#3\epsfxsize=#1 \epsfbox[0 0 4096 4096]{#5}
    \vskip -#1 \vskip #2 \vskip 10truept
    \vskip -0.2in
    \caption {#6 \label{#4}}
    \vskip 0.0truein plus0.2truein}
    \end{figure}
}
\def\BE{\begin{equation}}
\def\EE{\end{equation}}
\def\BEA{\begin{eqnarray}}
\def\BEAN{\begin{eqnarray*}}
\def\EEA{\end{eqnarray}}
\def\EEAN{\end{eqnarray*}}
\def\EL{\nonumber\\}

\newcommand{\op}{{\cal O}}
 
\title{Glueballs and Hybrids (Gluons as Constituents)}

\author{ Doug~Toussaint
\address{Department of Physics, University of Arizona, Tucson, AZ 85721, USA}
}

\begin{document}
\begin{abstract}
After a brief introduction to hybrid and glueball source operators, I
summarize recent lattice results for these particles.
\end{abstract}
\maketitle

%\section{Introduction}
\renewcommand{\thefootnote}{\fnsymbol{footnote}}

In addition to the familiar baryons made from three quarks and
mesons made from a quark and an antiquark, the spectrum of QCD is
expected to contain glueballs, hybrids and multiquark states.  Since the
earliest lattice QCD calculations it has been recognized that lattice
QCD should be used to calculate the properties of these particles.
Hopefully, lattice results will help to sort out which of the many
known hadrons contain important glueball or hybrid components.
The experimental situation for glueballs and hybrids with light
quarks was recently reviewed in Ref.~\cite{GodNap}.  Evidence for
an exotic $1^{-+}$ particle has been seen in several
experiments\cite{ExpExotic}, and is now substantial enough to merit
inclusion in the Review of Particle Properties\cite{PDG}.  Unfortunately, the
mass of this particle, 1400 MeV, is quite a bit lower than the lattice
results so far, and its decays conflict with phenomenological flux-tube
models.

Of course, we would like to know everything about these particles ---
their masses, their decays, their wavefunctions, their mixings \ldots .
However, the place to start, and where most of the calculations have
stopped, is with a calculation of the masses.  This works just like
spectrum calculations for ordinary hadrons: find an operator with
the desired quantum numbers $\op_i(\vec x,t)$, and compute
\BEA
&&\int d^3 x \langle \op_i(0) \op_j(\vec x,t) \rangle = \EL
&&A_i^{(0)}A_j^{(0)} e^{-m_0 t} + A_i^{(1)}A_j^{(1)} e^{-m_1 t} + \ldots\ \ \ .
\EEA
The ground state is the easiest to find; for excited states you
must manipulate several $\op_i$ and look for combinations with $A_i^{(0)}=0$,
or simultaneously fit to several masses.  Even for the ground state, it
is still important to construct an operator minimizing overlap with excited
states.

To make these operators, we can combine the following ingredients:\\
\rule{0.0in}{0.1in}\ \ $q^a$: quark, color $3$,\\
\rule{0.0in}{0.1in}\ \ $\bar q^a$: antiquark, color $\bar 3$, \\
\rule{0.0in}{0.1in}\ \ $\bar q^a \Gamma q^a$: color singlet quark bilinear\\
\rule{0.0in}{0.1in}\ \ $\bar q^a \Gamma q^b$: color octet quark bilinear\\
\rule{0.0in}{0.1in}\ \ $B^{ab}$: color magnetic field, color $8$,
$J^{PC}=1^{+-}$\\   
\rule{0.0in}{0.1in}\ \ $E^{ab}$: color electric field, color $8$,
$J^{PC}=1^{--}$\\
\rule{0.0in}{0.1in}\ \ $\vec L$: orbital angular momentum\\
For example, to make the simplest glueballs you might use:\\
\rule{0.0in}{0.1in}\ \ $B_i^{ab} B_i^{ab} = 0^{++}$ \\
\rule{0.0in}{0.1in}\ \ $B_i^{ab} B_j^{ab} - \frac{1}{3} Trace() = 2^{++}$ \\
\rule{0.0in}{0.1in\rule{0.0in}{0.1in}}\, \ $B_i^{ab} E_i^{ab} = 0^{-+}$ \\

\begin{figure}[t]
\rule{0.0in}{0.3in}\\
\epsfxsize=2.65in
\epsfbox[0 0 4096 4096]{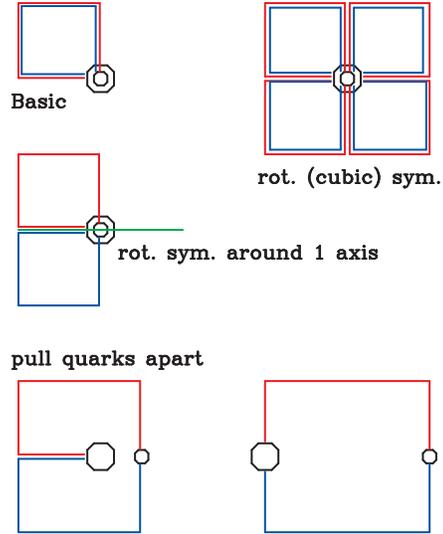} %}\\
%\epsfbox[2548 0 4596 2048]{hybridops.ps} %}\\
\rule{0.0in}{0.01in}\vspace{-0.5in}\\
\caption{
Construction of hybrid operators, illustrating the connection
between the ``local'' operators and ``excited potential'' operators.
\label{CARTOON}
}
\end{figure}

One way to understand the construction of the simplest hybrid
operators is to replace one of the glue fields in these glueball
operators with a $\bar q q$ bilinear.  In particular, take
a quark and in the 1S state, ``break'' the color, and insert
the color magnetic field.  In its simplest form, called ``Basic'' in the
upper left of Fig.~\ref{CARTOON},
the color magnetic field is just the difference between
the product of links in a small loop traversed clockwise and
counterclockwise connecting the quark spinor to the antiquark.

%\rule{0.0in}{0.1in}\\
%\epsfxsize=3.0in
%%\epsfbox[0 0 4096 4096]{colorbreak.ps} %}\\
%\epsfbox[2548 0 4596 2048]{colorbreak.ps} %}\\
%\rule{0.0in}{0.1in}\\
%{ Fig. XXX.  A hybrid operator, formed from a quark and antiquark
%(octagons) with their color indices linked by $F_{\mu\nu}$.
%}

The very
first thing you might try is to take a pion operator, 
$\bar q^a \gamma_5 q^a$, and combine it with the color magnetic
field to get $\bar q^a \gamma_5 q^b B_i^{ab}$.
The quantum numbers of this operator are found by combining the
quantum numbers of the pion and the magnetic field ---
 $0^{-+} \otimes 1^{+-} = 1^{--}$ --- the same quantum numbers as
the $\rho$.

Similarly, if you take the quark and antiquark to have spin one,
$\bar q^a \vec \gamma q^a$,
you can combine it with the color magnetic field to produce three
different quantum numbers:
$1^{--} \otimes 1^{+-} = 0^{-+} \oplus 1^{-+} \oplus 2^{-+}$\\
\ \ $0^{-+}$: $\bar q^a \gamma_i  q^b B_i^{ab}$: (q. num. of pion) \\ 
\ \ $1^{-+}$: $\epsilon_{ijk} \bar q^a \gamma_i  q^b B_j^{ab}$:
({\bf exotic }) \\
\ \ $2^{-+}$: $\bar q^a \gamma_i  q^b B_j^{ab} + (i\leftrightarrow j) -
Tr()$: (quantum  numbers of $\pi_2$)\\

Thus, if hybrid spectroscopy works like conventional meson spectroscopy
or glueball spectroscopy, we expect that the lowest multiplet of
hybrid states contains  $1^{--}$, $0^{-+}$, $\bf 1^{-+}$ and
$2^{-+}$ particles, both isovector and isoscalar, which will be
split by the color hyperfine interactions.  The most interesting of
these is the $1^{-+}$, which has exotic quantum numbers.

If we continue, and consider $\bar q q$ operators
with the quark and antiquark in a P wave in the nonrelativistic
limit, we can construct operators including exotics with
quantum numbers $0^{+-}$ and $2^{+-}$, and 
by using the color electric field you can make a $0^{--}$ exotic
($a_1 \otimes \vec E = 1^{++} \otimes 1^{--} = {\bf 0^{--}} \ldots$).

The ``Basic'' cartoon of a hybrid operator in Fig.~\ref{CARTOON}
is not really acceptable
since it violates the cubic symmetries of the lattice --- there is
nothing special about the upper left plaquette.  An obvious way to
symmetrize it is to average over all the nearby plaquettes, producing
the ``clover'' in the upper right of Fig.~\ref{CARTOON}.  Another way to proceed
is to begin by moving the traversal of the plaquette in one of the
two directions down to the lower left, as in the center of Fig.~\ref{CARTOON}.
This doesn't give the full cubic symmetry, but (with a few more loops
out of the plane of the figure) does have rotational symmetry around
one axis.  Then one can imagine pulling the quark and antiquark apart
along this axis, reaching the cartoon in the lower right of Fig.~\ref{CARTOON}
by way of the intermediate stage in the lower left.
In this way we make contact with a picture of a hybrid meson as an
excited flux tube connecting the quark and antiquark.

For both glueballs and hybrids it is necessary to use ``smeared''
or ``fuzzed'' links for the gluon parts of the operator to get good
overlap with the ground state particles.  The details of how this
%is done vary 
%from calculation to calculation
is done vary, and more often than not
the parameters are determined empirically.

Along with the choice of the operator which creates the hybrid,
we can choose how the quark and antiquark propagate.  The ideal way
is to use the full relativistic formulation.  However, this is time
consuming, and for heavy enough quarks it is much more sensible
to use the nonrelativistic (NRQCD) propagators.  In fact, one can
go all the way to the infinite mass limit, and simply use static
quarks.  (This was the first method used to study hybrids on the
lattice).  In addition to its simplicity, the use of static quarks
gives a very physical picture of the hybrid meson as a quark and
antiquark moving in a potential that depends on the state of the flux
tube connecting them (keywords: adiabatic potential, Born-Oppenheimer
approximation), and within this approximation it is straightforward
to compute the wavefunctions of the hybrids, as well
as masses of excited states.  (Since the quarks can't move in the
static approximation, it is necessary to use the ``flux tube'' versions
of the operators in Fig.~\ref{CARTOON}
so that the quark and antiquark can be
created at the desired separation.)

%SKIPPING DISTINCTION BETWEEN LATTICE AND CONTINUUM SYMMETRIES - OUT OF
%SPACE???  FIXXX

While the first lattice calculation of hybrids was done in 1983, 
progress was slow until the last few years.  Highlights of these
calculations can be found in Refs.~\cite{HY01} through \cite{HY16}.
%WHICH METHODS? FIXXX

%SPATIAL SIZE AND TIME RESOLUTION PROBLEMS. FIXXX

All of the lattice studies agree with our expectation that the $1^{-+}$
should be the lightest exotic hybrid, and this particle has received the
most attention.  To give a flavor for the state of the art, Table~\ref{HTABLE}
shows a selection of results for the mass of the $1^{-+}$ hybrid or, in
the case of heavy quarks, for the mass difference between the $1^{-+}$
hybrid and the $\bar q q$ S-wave mass.
The tabulated results show a pleasing convergence with time,
and for recent results (bold face)
the agreement among different groups is quite satisfactory.

Glueball mass calculations have a longer history than hybrid mass calculations.
Most have been done in the quenched approximation, perhaps
partly because glueballs are the actual excitations of quarkless QCD.
The state of the art is represented by three accurate calculations
in the quenched approximation: a UKQCD calculation\cite{UKQCDGLUE},
a series of works by the GF11 group\cite{GF11GLUE}
and a recent calculation by Morningstar and Peardon\cite{MPGLUE}.
The first two of these are conventional calculations, using isotropic
lattices with the one-plaquette action, while the third uses anisotropic
lattices with an improved action in the spatial planes.
All three 
%of these calculations 
use results at several lattice spacings
to get good control of the $a \rightarrow 0$ limit for the lowest glueballs.  
In addition, the anisotropic calculation\cite{MPGLUE} gets convincing
results for a few of the excited states.
As emphasized by Teper\cite{TEPERGLUE}, the UKQCD and GF11
calculations of the $0^{++}$ glueball mass in units of
the lattice spacing are in excellent agreement,
and the difference in their reported
results for the mass in MeV comes from different methods of assigning
a lattice spacing.  Some of this ambiguity is inherent in quenched
calculations, since we expect that quenched mass ratios in the
$a\rightarrow 0$ limit will differ from real world mass ratios.
To compare these calculations I follow contemporary practice and use
$r_0$ to set the lattice spacing.  Ref.~\cite{MPGLUE} quotes masses
in units of $r_0$, and for the conventional action calculations
I convert to physical units using the interpolating
formula for $r_0$ of Guagnelli, Sommer and Witting\cite{GUAGNELLI}
%and declaring that $r_0=0.50$ fm.  With this definition of lattice
using $r_0=0.50$ fm.  With this definition of lattice
spacing, some of the results of these three calculations are:

\begin{tabular}{ll}
\multicolumn{2}{l}{ UKQCD } \\
\ \ $0^{++}$ & 1645(50) MeV\\ %XXX CHECK THIS
\ \ $2^{++}$ & 2337(100) MeV\\
\multicolumn{2}{l}{ GF11 } \\
\ \ $0^{++}$ & 1686(24)(10) MeV\\
\ \ $2^{++}$ & 2380(67)(14) MeV\\
\multicolumn{2}{l}{ Morningstar and Peardon } \\
\ \ $0^{++}$ & 1659(43)(16) MeV\\
%\ \ $2^{++}$ & 2561(173)(28) MeV\\ ERROR (caught by Urs)
\ \ $2^{++}$ & 2304(8)(24) MeV\\
\ \ $0^{-+}$ & 2494(28)(24) MeV\\
\ \ $0^{++*}$ & 2561(173)(28) MeV\\
\ \ $2^{++*}$ & 3499(43)(35) MeV\\
\end{tabular}
% beta=6.4, a/r0 = .1025

All three calculations are in reasonable agreement for the $0^{++}$
and $2^{++}$ glueball masses, and the Morningstar and Peardon results
show the state of the art in extracting excited state masses.

There is also a growing body of work, much of it by the GF11
group\cite{GF11GLUE}, on the decay rates of glueballs, and on the mixing
of glueballs with quarkonia to produce the physically observed hadrons.
Here the idea is to take the quenched glueball masses and the quenched
light quark isoscalar and $\bar s s$ quarkonia masses, together with
either a measurement or a model for their overlap, and diagonalize a
matrix to find the physical states.  (Oddly, this is a calculation where
we actually prefer a quenched mass spectrum to a full QCD lattice
calculation.)  The eigenstates are then identified with the experimental
$f_0(1370)$, $f_0(1500)$ and $f_J(1710)$ (assuming $J=0$).
Such an analysis has been done by the GF11 group using their own
numerical results\cite{GF11GLUE} and by Teper using numerical inputs from the
literature\cite{TEPERGLUE}.

Teper proposes:
\BEAN
f(1710) &=& 0.42 g + 0.90 \bar ss + 0.13 \bar uu \EL
f(1500) &=& 0.77 g - 0.43 \bar ss + 0.48 \bar uu \EL
f(1370) &=& -0.49 g + 0.10 \bar ss + 0.87 \bar uu 
\EEAN
while the GF11 group suggests:
\BEAN
f(1710) &=& 0.86(5) g+0.30(5) \bar ss +0.41(9) \bar uu \EL
f(1500) &=& -0.13(5) g+0.91(4) \bar ss -0.40(11) \bar uu \EL
f(1370) &=& -0.50(12) g+0.29(9) \bar ss +0.82(9) \bar uu \EL
\EEAN
\rule{0.0in}{0.1in}\vspace{-0.3in}\\
where $g$ is the glueball component, $\bar s s$ the strangeonium
component,\pagebreak and $\bar u u$ the (isoscalar) light quark component.

These results are quite different.
The largest difference comes from differences in the unmixed masses,
with Teper having 
$m_{\bar u u}=1.36$, $m_{\bar s s}=1.61$ and $m_{g}=1.48$ GeV
and the GF11 group having
$m_{\bar u u}=1.47$, $m_{\bar s s}=1.51$ and $m_{g}=1.63$ GeV.
Note that in the first analysis the unmixed $\bar s s$ is heavier than
the glueball, while in the second it is lighter.
Not surprisingly, the heaviest physical particle is mostly $\bar s s$ or
mostly glueball, respectively.  Clearly it is important to know whether
the quenched $\bar s s$ is heavier than the quenched glueball, and the
%GF11 group has attacked this question by calculating quarkonium masses
%on the same samples used in the glueball calculations.
GF11 group has attacked this question by calculating quarkonium 
and glueball masses on the same samples.

%Teper
% f(1710) = 0.42g + 0.90ss + 0.13 uu
% f(1500) = 0.77g - 0.43ss + 0.48 uu
% f(1370) = -0.49g + 0.10ss + 0.87uu
%\put(0.7,-0.3){\color{Green}$\bar u u$}
%\put(1.4,-0.3){\color{Red}$\bar s s$}
%\put(2.1,-0.3){\color{Blue}$g g$}
%\put(0.5,-0.7){1.36}
%\put(1.3,-0.7){1.61}
%\put(2.1,-0.7){1.48}
%Weingarten
% f(1710) = .859(54)g+0.302(52)ss+0.413(87)uu
% f(1500) = -0.128(52)g+0.908(37)ss-0.399(113)uu
% f(1370) = -0.495(118)g+0.290(91)ss+0.819(89)uu
%\put(3.2,-0.3){\color{Green}$\bar u u$}
%\put(3.9,-0.3){\color{Red}$\bar s s$}
%\put(4.6,-0.3){\color{Blue}$g g$}
%\put(3.0,-0.7){1.47}
%\put(3.8,-0.7){1.51}
%\put(4.6,-0.7){1.63}

\begin{table}[th]
\caption{
{\bf Some results for $\bf 1^{-+}$ hybrid masses}
\label{HTABLE}
}
\begin{tabular}{lllll}
Date	& Ref.	& Method	& $\Delta M$ (GeV) &  \\
\multicolumn{5}{l}{ \large\bf $\bar b b g - \bar b b$: } \\
1990	& \cite{HY06} & St.	& 1.11(3)(?) 	&		\\
1993	& \cite{HY07} & NR.		& 0.8(?)(?) 	&		\\
1997	& \cite{HY10} & NR.	& 1.68(10) 	&		\\
	&	      &		& 1.40(14) 	& (1)	\\
1997	& \cite{HY12} & NR.	& 1.14(21) 	&		\\
1997	& \cite{HY11} & St.	& 1.3		&		\\
1998	& \cite{HY15} & NR.(An.)	& {\bf 1.542(8)}	& \\
1999	& \cite{HY16} &St+NR(An.)		& {\bf 1.49(2)(5)}	& \\
\multicolumn{5}{l}{ \large\bf $\bar c c g - \bar c c$: } \\
1990	& \cite{HY06} & St.	& 0.94(3)	&		\\
1996	& \cite{HY09} & Rel.(Wil.)		& {\bf 1.34(8)(20)}	& \\
1998	& \cite{HY14} & Rel.(Clo.)		& {\bf 1.22(15)(?)}	& \\
1999	& \cite{HY15} & NR.(An.)	& {\bf 1.323(13)}	& \\
\multicolumn{3}{l}{ \large\bf $\bar s s g$ } & M (GeV.) &  \\
1996	& \cite{HY08} & Rel.(Clo.) & {\bf 2.00(20)}	& (2)\\
1996	& \cite{HY09} & Rel.(Wil.) & {\bf 2.17(8)(20)}	& (3)\\
\multicolumn{5}{l}{ \large\bf $\bar u d g$ } \\
1996	& \cite{HY08} & Rel.(Clo.) & {\bf 1.88(20)}	& (4)\\
1996	& \cite{HY09} & Rel.(Wil.) & {\bf 1.97(9)(30)}	& (5)\\
1998	& \cite{HY13} & Rel.(Wil.) & {\bf 1.90(20)}	& (6)\\
1998	& \cite{HY14} & Rel.(Clo.) & {\bf 2.11(10)(?)}	& (7)\\
\end{tabular}
{Abbreviations:
St. $\!=\!$ Static,
NR. $\!=\!$ NRQCD,
Rel. $\!=\!$ Relativistic,
An. $\!=\!$ anisotropic,
Wil. $\!=\!$ Wilson,
Clo. $\!=\!$ clover.\\
Notes:
(1): value with $a$ determined differently,
(2): $a=0.095$ fm,
(3): $a=0.075$ fm,
(4): Model to extrapolate to $m_q=0$, 120 MeV below $\bar s s$ mass, $a=0.095$ fm,  
(5): Extrapolation from several $am_q$ values, $a=0.075$ fm,
(6): $N_f=2$ dynamical quarks, extrapolate from several $am_q$ values, $a=0.086$ fm,
(7): Same as (5).
\rule{0.0in}{0.01in}\vspace{-0.5in}\\
}
\end{table}

It is also interesting to test the effects of dynamical quarks on the
glueball spectrum.  In principle this is tricky, since a full QCD
spectrum calculation with a glueball source operator will produce the
masses of the physical states, which are mixtures of glueballs and
quarkonia.  In practice, the calculations that have been done so far
have used quark masses large enough that the quarkonium mass exceeds the
glueball mass, and it is reasonable to simply say thay they have
measured the glueball mass.  The largest calculation is from the
T$\chi$L/SESAM collaborations\cite{SESAMGLUE}, and another preliminary result
from UKQCD was presented at this conference\cite{MICHAELGLUE}.
The T$\chi$L/SESAM results for the $0^{++}$ glueball
 are within errors of the quenched results,
although they do see hints of a larger dependence on the lattice size.
However, the preliminary UKQCD results have a much smaller mass, despite
being done at approximately the same sea quark mass, as measured by
$m_\pi/m_\rho$.  The UKQCD results are done on much coarser lattices,
albeit with the clover action rather than the Wilson quark action.
(UKQCD used $\beta=5.2$ while T$\chi$L/SESAM used 5.6.)
The $0^{++}$ glueball mass has long been known to be small on coarse
lattices\cite{SMALLSCALARGLUE},
but the preliminary UKQCD results are even smaller than we
would expect from our experience with the quenched theory.
%INCLUDE THE FIGURE??? FIXXX

Since the exotic $1^{-+}$ signal found in experiments at 1400 MeV is
much lower than expected from lattice calculations (and most other
theoretical approaches), it is tempting to ask whether it could be
something else, most likely a 4-quark ($\bar q \bar q q q$) state.
%Such states, as well as the much discussed H-dibaryon ($qqqqqq$),
%represent another way in which full QCD might be richer than the simple
%quark model.
In principle, this question is answerable with lattice methods, but it
is a difficult subject.  Nonetheless, there is a small but growing body
of work on lattice 4-quark states\cite{FOURQUARK,FOURQUARK2},
beginning with the
simplest case where all four quarks are static, and moving into the case
of two static and two moving quarks.  An amusing limit has recently
been studied by Michael and Pennanen\cite{FOURQUARK2}, where the two
quarks are very heavy, and the two antiquarks light.  The two quarks
have an attractive interaction in the $\bar 3$ color combination, and
since they are very heavy they can bind into a small object which is
effectively an antiquark.  Then the three antiquarks may be regarded
as an antibaryon, with one constituent very heavy.

I expect that lattice calculations will continue to be important in
sorting out the rich structure of the QCD spectrum.  It is important to
extend our work to calculations of decay rates and mixings among
particles.

\section*{Acknowledgements}
I am grateful to the organizers of Lattice-99 for the
opportunity to present this talk.
I thank Gunnar Bali, Julius Kuti,                       
Thomas Manke, Craig McNeile, Chris Michael, Colin Morningstar,
Petrus Pennanen, Mike Teper
and Don Weingarten for providing results used in this talk.
This work was supported by the U.S. 
%DOE.
Department of Energy.


\begin{thebibliography}{99}

%EXPERIMENT
\bibitem{GodNap}
S.~Godfrey~and~J.~Napolitano, hep-ph/9811410, submitted
to Rev. Mod. Phys.

\bibitem{ExpExotic}
D.R. Thompson {\it et al.}, Phys. Rev. Lett. {\bf 79} (1997) 1630;
A. Abele {\it et al.}, Phys. Lett. {\bf B423} (1998) 175.

\bibitem{PDG}
%Review of Particle Properties,
European Physical Journal C, {\bf 3} (1998).

% HYBRID BIBLIOGRAPHY
\bibitem{HY01} L.A. Griffiths, C. Michael and P.E.L. Rakow,
Phys. Lett. {\bf 129B} (1983) 351.

\bibitem{HY02} J.E. Mandula,
Phys. Lett. {\bf 135} (1984) 155.

\bibitem{HY03} N.A. Campbell, C. Michael and P.E.L. Rakow,
Phys. Lett. {\bf 139B} (1984) 288.

\bibitem{HY04} N.A. Campbell {\it et al.},
Phys. Lett. {\bf 142B} (1984) 291.

\bibitem{HY05} N.A. Campbell, A. Huntley and C. Michael,
Nucl. Phys. B {\bf 306} (1988) 51.

\bibitem{HY06} S. Perantonis and C. Michael,
Nucl. Phys. B {\bf 347} (1990) 854.

\bibitem{HY07} S.M. Catterall, F.R. Devlin,
I.T. Drummond and R.R.  Horgan, %UKQCD
Phys. Lett. {\bf 300B} (1993) 393.

\bibitem{HY08} P. Lacock, C. Michael, P. Boyle and P. Rowland, %UKQCD
%hep-lat/9605025,
Phys. Rev. D {\bf 54} (1996) 6997;
%hep-lat/9611011,
Phys. Lett. {\bf 401B} (1997) 308.

\bibitem{HY09} C. Bernard {\it et al.} (MILC),
%hep-lat/9607031,
Nucl. Phys. (Proc. Suppl.) {\bf 53} (1997) 228;
%hep-lat/9707008,
Phys. Rev. D {\bf 56} (1997) 7039.

\bibitem{HY10} T. Manke, I.T. Drummond, R.R. Horgan and H.P. Shanahan, %(UKQCD),
%hep-lat/9710083,
Phys. Rev. D {\bf 57} (1998) 3829;
%hep-lat/9709001,
Nucl. Phys. (Proc. Suppl.) {\bf 63} (1998) 332.

\bibitem{HY11} K.J. Juge, J. Kuti and C.J. Morningstar,
%hep-lat/9709131,
Nucl. Phys. (Proc. Suppl.) {\bf 63} (1998) 326,
hep-lat/9809015.

\bibitem{HY12} S. Collins, C. Davies and G. Bali, % (UKQCD),
%hep-lat/9710058,
Nucl. Phys. (Proc. Suppl.) {\bf 63} (1998) 335.

\bibitem{HY13} P. Lacock and K. Schilling, % (SESAM),
%hep-lat/9809022,
Nucl. Phys. (Proc. Suppl.) {\bf 73} (1999) 261.

\bibitem{HY14} C. Bernard {\it et al.} (MILC),
%hep-lat/9809087,
Nucl. Phys. (Proc. Suppl.) {\bf 73} (1999) 264.

\bibitem{HY15} T. Manke {\it et al.} (CP-PACS),
%hep-lat/9812017,
Phys. Rev. Lett. {\bf 82} (1999) 4396.

\bibitem{HY16} K.J. Juge, J. Kuti and C.J. Morningstar,
%hep-ph/9902336,
Phys. Rev. Lett. {\bf 82} (1999) 4400.

% DISTINGUISH J=0 AND J=4
%\bibitem{XXX}
%M.Teper,
%Nucl. Phys. (Proc. Suppl.) {\bf 73} (1999) 264.

%GLUEBALLS
% THREE ACCURATE QUENCHED CALCS

\bibitem{UKQCDGLUE}
G. Bali {\it et al.} %(UKQCD)
%hep-lat/9304012,
Phys. Lett. {\bf B309} (1993) 378.

\bibitem{GF11GLUE}
H. Chen, J. Sexton, A. Vaccarino and D. Weingarten,
hep-lat/9308010, %XXX
%hep-lat/XXX,
Nucl. Phys. B (Proc. Suppl.) {\bf 34}, 357 (1993);
W. Lee and D. Weingarten,
hep-lat 9805029,
% Scalar Quarkonium Masses and Mixing with the Lightest Scalar Glueball
%hep-lat/9801013,
Nucl. Phys. (Proc. Suppl.) {\bf 63} (1998) 194;
%The Scalar Quarkonium Spectrum and Quarkonium-Glueball Mixing
A. Vaccarino and D. Weingarten, in preparation.
%glueball mass predictions ... (preliminary)

\bibitem{MPGLUE}
C.J. Morningstar and M. Peardon  %anisotropic lattices
%hep-lat/9704011,
Phys. Rev. D {\bf 56} (1997) 4043,
%Efficient glueball simulations on anisotropic lattices
hep-lat/9901004.
%Phys. Rev. D ???
% The glueball spectrum from an anisotropic lattice study

\bibitem{TEPERGLUE}
M. Teper, % (reviews)
hep-lat/9711011, %Physics from the lattice: glueballs in QCD; topology;
SU(N) for all N
hep-ph/9711299, %What lattice calculations tell us about the glueball spectrum
hep-th/9812187.

\bibitem{GUAGNELLI}
M.~Gaugnelli, R.~Sommer and H.~Wittig,
Nucl. Phys. {\bf B535} (1998) 389.

%DYNAMICAL

\bibitem{SESAMGLUE}
%hep-lat/9608096,
G. Bali {\it et al.} Nucl. Phys. (Proc. Suppl.) {\bf 53} (1997) 239;
%hep-lat/9710012
Nucl. Phys. (Proc. Suppl.) {\bf 63} (1998) 209;
G.~Bali, private communication.

\bibitem{MICHAELGLUE}
C. Michael, these proceedings.

\bibitem{SMALLSCALARGLUE}
A.~Patel {\it et al.}, Phys. Rev. Lett. {\bf 57} (1986) 1288.

%FOUR QUARK STATES
\bibitem{FOURQUARK}
Y. Liang, B. A. Li, K.F. liu, T. Draper,  and R.M. Woloshyn,
Nucl. Phys. B (Proc. Suppl.) {\bf 17} (1990) 408; % rho-rho states
A.M. Green, C. Michael and J.E. Paton, Nucl. Phys. {\bf A554} (1993) 701;
A.M. Green, C. Michael, J.E. Paton and M.E. Sainio,
Int. J. Mod. Phys. {\bf E2} (1993) 479;
S. Furui, A.M. Green and B. Masud, %hep-lat/9409006;
Nucl. Phys. {\bf A582} (1995) 682;
A.M. Green, J. Lukkarinen, P. Pennanen and C. Michael, %hep-lat/9508002;
Phys.Rev. D {\bf 53} (1996) 261;
P. Pennanen, %hep-lat/9608147;
Phys.Rev. D {\bf 55} (1997) 3958.

\bibitem{FOURQUARK2}
C. Michael and P. Pennanen, %hep-lat/9901007.
Phys.Rev. D {\bf 60} (1999) 054012;
these proceedings.

\end{thebibliography}
\end{document}